\documentclass[12pt]{article}
\begin{document}

\title{The Anthropic Principle Revisited}

\author{Berndt M\"uller\\
        Department of Physics, Duke University\\
        Durham, NC 27708-0305, USA}
\date{}
\maketitle

\abstract{The usage of the anthropic principle in modern cosmology
is reviewed. It is argued that its recent use to explain the observed 
values of cosmological parameters as most probable values for an 
ensemble of universes, is not justified. However, the anthropic 
principle can be invoked to argue that a vast number of universes 
must exist, in which the cosmological parameters and constants of 
nature take on different values. This argument lends support to the 
hypothesis of eternal inflation and suggests that at least some of 
the parameters of the Standard Model cannot be quantitatively derived
from an underlying theory.}

\section{Introduction}

The anthropic principle\cite{Ca74,CR79,Dy79,Da82,BT86,JG00} occupies 
a peculiar place in the arsenal of scientific reasoning. On the one 
hand, physicists generally harbor a deeply held belief that the 
fundamental laws of nature have an objective origin and are 
not human constructs or random accidents. On the other hand, 
some numerical coincidences in the natural laws appear far
too contrived to be easily reconciled with our sense of 
``naturalness''. The anthropic principle is concerned with 
those fortuitous coincidences that have relevance for our own 
existence.

There are many well-known cases of such coincidences. As one 
example, consider the resonance in the excitation spectrum of 
$^{12}$C at 7.65 MeV. The precise location of this state 
is of critical importance for the fusion of three $^4$He nuclei 
into a $^{12}$C nucleus, and thus is a determining factor for the 
abundance of carbon and heavier atoms in the universe. Indeed,
the presence of this resonance is so essential, that its existence
was predicted by Hoyle\cite{Ho54} in order to explain the observed 
pattern of stellar nucleosynthesis. Obviously, the position of this 
state depends in extremely subtle ways on the strength of the nuclear
force which determines the structure of the $^{12}$C nucleus. It has
been estimated\cite{OPC99} that almost no carbon would be synthesized
in stars if the nucleon-nucleon interaction were weaker by just 4\%.
Several fundamental physical constants combine to determine the
strength of the effective nuclear interaction; if considered as
a function of the Higgs vacuum expectation value $v$ alone, the
anthropic principle requires\cite{JS00} that $v/v_{\rm obs}>0.9$.

Other well-known instances of apparent fine-tuning of fundamental 
physical constants\cite{Da82,PL83,ABDS} are the neutron-proton mass 
difference, and the relation ($G$ is the gravitational constant, 
$m_e$ and $m_p$ are the electron and proton mass, respectively, 
and $\alpha$ denotes the electromagnetic coupling constant)
\begin{equation}
G m_p^2/\hbar c \approx 3 \alpha^{12} (m_e/m_p)^4 
\approx 6\times 10^{-39} ,
\end{equation}
which is an essential prerequisite for the existence of hydrogen
burning stars with similar mass and luminosity as our sun.

Of course, if intelligent life had not developed anywhere in the 
universe, no one would be around to worry about those problems. 
But we do exist and hence face the challenge of having to explain 
this fact. The question I address in this article is, in which terms 
the discussion should be framed within the rules of scientific 
reasoning and common sense. I first review some of the traditional 
formulations of the anthropic principle (in Section 2) and describe 
its usage in modern cosmology (in Section 3). In Section 4, I present 
my own, somewhat different, view of the subject.

\section{Forms of the Anthropic Principle}

In its weakest form, the anthropic principle states that the laws of 
nature and the evolution of our universe must be compatible with the 
fact that we exist. In this form, the anthropic principle is almost 
a tautology. It might be used, in this weak form, to invalidate an 
experimental result, e.~g.~if it could be rigorously established that 
the result would imply our nonexistence. However, such a reasoning is 
unlikely to be of great usefulness, because we do not have a firm 
grasp of all the requirements for the formation of intelligent life 
in the universe. In its weak form the anthropic principle also serves
to remind us that certain aspects of the history of our universe may
be biased by the outcome, i.~e.~the fact of our existence.

In a stronger form, the anthropic principle is understood to explain 
(if they have already been measured) or to predict (if they are not 
yet known) certain aspects of nature, such as numerical values of
physical constants or the properties of atoms and nuclei. The
prediction of the fusion resonance in $^{12}$C is a famous example
of such an application of the anthropic principle. Another recently 
much debated example is the ``unnaturally'' small value of the 
cosmological constant\cite{CCrefs}
\begin{equation}
\Lambda \approx (2\times 10^{-3} {\rm eV})^4 .
\label{CC}
\end{equation}
If $\Lambda$ were larger than the observed value by a factor
200 or more, the universe could not have gone through the slow 
period of expansion that is required for the formation of large
galaxies, and it is unlikely that life could have developed\cite{We87}. 
On the other hand, the ``naturally'' expected value 
of the cosmological constant as the vacuum energy in the standard 
(supersymmetric) model of particle physics is at least 60 orders 
of magnitude larger than the observed value\cite{SUSY}. The 
anthropic principle thus permits to select a much reduced range 
of expected values of the cosmological constant, even before its 
measurement.

In its strongest form, the anthropic principle has been used to 
explain, why the laws of nature and especially numerical constants 
in these laws, such as the coupling constants of fundamental forces
and the ratios of elementary particle masses, or the parameters of
cosmological models, have the specific forms or values that are 
observed. Of course, this use of the anthropic principle makes 
sense only where we lack other explanations. As the laws of
physics have become ever more unified and better understood, the 
number of truly fundamental, still unexplained parameters has been 
reduced to less than thirty.

\section{Inflationary Cosmology}

The philosophical basis of the anthropic principle has been 
radically changed by the inflationary model of the 
universe\cite{Gu81,Li82,AS82}. 
The hypothesis, that our universe went through a period of 
exponential growth about 14 billion years ago, provides a 
natural explanation for several astounding astrophysical facts. 
In particular, the inflationary model can explain the observed 
isotropy and homogeneity of the cosmic background radiation, 
the vast size of the universe compared with the fundamental
scales of particle physics, and the near flatness of the 
geometry of the universe on large scales\cite{Gu00}.

Cosmic inflation occurs naturally when the some region of 
space-time gets trapped in an excited vacuum state, usually
called the ``false'' vacuum\cite{KOV74,Fr76,Co77}.
The vacuum energy density and pressure induce an effectively 
repulsive gravitational force, which dictates the expansion of 
space-time. As long as the false vacuum persists the 
universe expands, doubling its volume in equal periods of time. 
When the false vacuum finally decays, the expansion slows down, 
the universe heats up to a very high temperature, and the familiar 
``Big Bang'' occurs. In order to explain the observations, the 
linear size of the universe must have grown by at least 26 orders
of magnitude, and possibly much more, during the inflationary
period.

However, the inflationary cosmological models have brought us 
another philosophically important and unanticipated insight. 
In most models, an infinite sequence of universes are created 
once inflation has started\cite{St83,Vi83}. The reason for this
behavior is that the volume of space filled by the excited vacuum 
state inflates faster than the excited state can decay. In other 
words, as the false vacuum decays in one bubble-like region of 
space, creating a new universe, the remaining space continues 
to grow exponentially, allowing for the formation of other bubbles 
in which the excited vacuum decays, and so on {\it ad infinitum}. 
This scenario is called ``eternal'' inflation.

The various universes created by this process are causally 
disconnected, and observers in different ones cannot communicate 
with each other. In the true sense of the word, eternal inflation
leads to a ``many worlds'' picture where an unlimited number of
big-bang universes develop in isolation, separated by rapidly
stretching regions of space filled with the undecayed false vacuum.
This has important consequences for the anthropic principle, because 
it is easy to construct models where certain ``constants'' of nature 
differ in value from one universe to the next. The anthropic principle 
in its weakest form will then simply assert that we live in one of 
those universes that are conducive to the formation of intelligent life.

A slightly different scenario is ``open inflation'', where each
created universe is composed of infinitely many regions with
different values of the cosmological parameters and possibly of
physical constants that depend on the vacuum structure
\cite{LLM94,GLL94,LM95}. Because each universe has an infinite 
volume, these regions can be so large that each extends beyond
the horizon of an observer living in it. Thus any observer would 
measure unique values for all cosmological and physical parameters 
throughout the entire range of the observable universe.

These cosmological scenarios invite novel applications of the 
anthropic principle,
because one can now, at least in principle, define the probability
distribution for various cosmological parameters and other vacuum
state dependent physical constants over the entire set of universes
\cite{LLM94,Vi95}. If this distribution can be calculated for a 
specific model, one can determine the most probable set of 
parameters and ask whether it permits the evolution of intelligent 
life. In other words, one postulates that we probably live in the 
most common type of universe or subuniverse. This postulate is 
sometimes called the {\it principle of mediocrity}\cite{Vi95}.

If we denote the set of physical parameters that can take different
values in different regions of space-time by $\chi$, the probability
for measuring a particular value at some space-time point will be
proportional to the space-time volume $V(\chi)$ of the region in 
which this value is realized. The probability that this value will
actually be measured by an intelligent observer will, furthermore,
be proportional to the number $\nu(\chi)$ of large galaxies formed
in the presence of this value of $\chi$, and the number $c(\chi)$
of advanced civilizations developing in each galaxy. Altogether, 
one has the following relation for the probability density that 
the value $\chi$ will be observed:
\begin{equation}
P_{\rm obs}(\chi) d\chi \sim c(\chi) \nu(\chi) V(\chi) d\chi .
\label{prob}
\end{equation}
The mediocrity principle then postulates that the value $\chi^*$
measured experimentally is near the maximum of $P_{\rm obs}(\chi)$.

For the case, in which each bubble universe is characterized by a
unique value of parameters, the definition of a probability density
encounters certain ambiguities due to the infinity of universes 
createdby the process of eternal inflation. These problems are 
connected with the ambiguities in the definition of a universal 
time coordinate\cite{LLM94,LM96}. In the case of open inflation, 
the definition of a unique probability measure is much simpler, 
because all possible values of the parameters are realized in the 
subregions of each universe, and the volume of each homogeneous 
region at a fixed cosmological time can be used as a basis for 
the probability measure\cite{Vi98}. 

Applications of this concept to the cosmological constant problem 
are able to provide a possible explanation for the smallness of 
the observed value of the cosmological constant\cite{MSW98,We00}, 
as well as for its close coincidence at present with the total 
matter density in our universe\cite{GLV00,Bl00,Vi01,Do01}. 
However, it has been pointed out that the anthropic argument 
loses some of its predictive power, when not just isolated
parameter changes are considered, but correlated changes in the 
values of several cosmological parameters are considered
simultaneously\cite{TR98,Ag01}. It then appears possible to 
construct cosmologies with widely differing values of their 
parameters, which allow for the formation of sun-like stars 
with planetary systems.

\section{A Modified Form of the Anthropic Principle}

Now I want to address the question whether the ``principle of
mediocrity'', i.~e.~the notion that we live in an average, or most 
probable, universe among all possible universes, constitutes a 
logically compelling use of the anthropic principle. I will argue 
that it does not, and that the anthropic principle should instead 
be applied in a slightly different, but still very powerful way.

\subsection{There's no Reason for Being ``Average''}

The principle of mediocrity is the complete opposite of the idea
of geocentrism, which fell into disrepute after the successive 
discoveries of the heliocentric structure of the solar system, 
the great extent of our galaxy, and the vast number of galaxies 
in the visible universe. In fact, modern cosmology has entirely 
done away with the concept of a central point in the universe, 
rendering all locations equivalent, at least on very large scales. 
However, the knowledge that any two points in our universe are
equivalent, if viewed on a very large scale, does not imply that
the Earth looks like the ``average'' planet that harbors life. 
As long as we have not detected a single other earth-like planet 
in our galaxy, let alone one inhabited by intelligent beings, 
we have no reason to believe that astronomers will eventually 
find that intelligent forms of life usually develop on planets
similar to ours.

It is important to recognize that the demise of the geocentric 
model was the result, not of logical reasoning, but of observations 
that rendered it obsolete. But we have no such observations as 
guidance concerning the multitude of universes and, almost certainly, 
will never have them. We are only assured of our existence and the 
fact that the values of the fundamental physical constants and 
cosmological parameters conspire to make our existence possible.

In order to better expose the fallacy of the mediocrity principle,
it is useful to consider the following question: Can we explain the
peculiar properties of the planet Earth and explain our existence
on this planet, given the physical laws that we know? Many 
fortuitous circumstances have made intelligent life possible on
this planet: The sun has just the right mass to radiate enough
energy to sustain life based on organic compounds and water, 
and its longevity as an active star is large enough to permit
higher life forms to evolve. Earth has the right size and just 
the right distance from the sun to retain large amounts of liquid 
water at its surface. And the solar system exhibited just the 
right intensity of asteroidal impact activity during earlier periods 
to drive the evolution of higher forms of life by producing moments 
of punctuated equilibrium, without wiping life totally out or 
suppressing anything but simple organisms. And so forth.

Because we do not yet understand the processes governing the
formation of solar systems in sufficient detail, we cannot reliably 
estimate the probability for the presence of a planet with
similar properties as the Earth in a galaxy like ours. We also
cannot calculate the probability for the emergence of life,
or even intelligent life, on a planet with these properties.
What we can say with a fair degree of certainty, however, is
that the probability for finding the right combination of
circumstances capable of creating intelligent life in the 
vicinity of a given star must be quite small\cite{Drake}.

Unless we want to claim that we are the product of 
unlikely coincidences, or of purposeful and clever design, we 
must therefore assume that the universe contains a great number 
of stars, so that the probability for us to exist somewhere in 
the universe is not small\cite{FR98}. This does not assert that 
the Earth looks like the {\em average} planet inhabited by 
intelligent life forms. There is no justification for this 
assumption. In view of the many fortuitous circumstances 
which facilitated the emergence of intelligent life on Earth 
-- just imagine how life might have developed if the asteroid
that wiped out the dinosaurs had missed the Earth's path --
it would not at all be surprising if another civilization were
found to inhabit a very different looking planet. But there is 
no need to make this assumption. All that we need to assume is 
that there are so many solar systems in the universe, that the 
probability for finding one on which intelligent life can and 
will develop is close to one\cite{FR98}. 

By virtue of this reasoning, an intelligent observer could have
anticipated the vast size of our universe and the multitude of
stars contained in it, even if the view outside our solar system
were shrouded by a thick cover of a interstellar dust.
Precisely the high degree of improbability of the emergence of
intelligent life on our planet would force this conclusion upon
the imagined observer, unless he or she would want to believe 
that somehow all the fortuitous events and circumstances were 
arranged purposefully by some higher power\cite{Da92}.

\subsection{The Case for a Multitude of Universes}

We are now ready to apply the same reasoning to the physical laws
of nature and our universe as a whole. Without doubt, most random
choices for the values of the fundamental physical constants (the
coupling strengths of fundamental interactions, the masses of the 
elementary particles, and the cosmological parameters) would not
have permitted anything even remotely resembling intelligent life
to emerge. Then universe could have recollapsed after a fraction of
a second, its expansion could have begun to accelerate long before
galaxies formed, a proton and a neutron might not form a bound state,
or the neutron might have a much shorter lifetime. 

Even if the dream of particle physics, the unification of all 
forces and the formulation of a ``Theory of Everything'', were 
fulfilled, one would have to wonder why we are blessed with this 
particular set of fundamental laws and cosmological parameters 
that facilitate the emergence of intelligent life. Could the 
formation of life, especially intelligent life, be the necessary, 
or even probable, consequence of the fact that there is unique 
set of laws governing all processes in our universe? I find this 
hard to believe, and there is no evidence supporting this 
hypothesis.

We are thus led to formulate a new version of the anthropic 
principle: 
\begin{itemize}
\item[] {\em The physical laws of nature and the values of the 
fundamental constants and cosmological parameters must allow for 
the probable emergence of intelligent life forms. The values of 
these constants must be probable ones.} 
\end{itemize}
In other words, we understand the anthropic principle to mean
that our own existence should not be the consequence of an 
unlikely coincidence nor the result of divine design. But note 
that it was precisely the recognition that our existence is 
predicated on various highly improbable coincidences, which 
brought the anthropic principle to the attention of scientists.

The simplest resolution of this conundrum is to postulate the
existence of many different universes, or of large unobservable
regions in our infinitely extended universe, where the physical 
laws and parameters differ from those observed by us. There must 
exist so many of them that the probability for finding one that 
allows intelligent life to form is not much less than one. This 
hypothesis implies that some of the physical constants, if not all, 
can take different values. It also implies that it is impossible 
to calculate and predict all physical constants from fundamental 
principles, because their values are not fundamentally unique. 
It may well be possible to relate them to some underlying 
geometric structure of space-time, such as the arrangement of 
D-branes in higher dimensions\cite{ADD98}, but then this 
arrangement must be -- to a certain extent -- random and 
unpredictable.

When one adopts this form of the anthropic principle, the fact 
that eternal inflation predicts just such a scenario, entailing 
the creation of infinitely many almost flat universes with 
quasi-random values of the fields that determine the values of 
the physical constants, becomes a highly desirable aspect of this 
cosmological theory. The fact that superstring theory seems to 
have great difficulty predicting a unique vacuum state, becomes 
a benefit, rather than a drawback of this theory, as well. In the 
framework of our version of the anthropic principle, these are 
actually aspects that one would demand of any sensible Theory of 
Everything.

\section{Summary}

I have argued that the proper resolution of the various unnatural
coincidences among values of physical constants that have permitted
intelligent life to develop in our universe is the hypothesis that 
there must exist so many universes with different values for these 
constants that, overall, the probability of finding one with 
favorable conditions is reasonably close to one. This postulate
has several far-reaching implications. It dictates that not all
constants of nature are uniquely predictable, that some must be able 
to take a range of random values. It demands that our observable
universe is only one among many in existence, and there is no
reason for the assumption that our universe has highly probable
characteristics within this ensemble. 

It is quite remarkable that many modern cosmological models and
unified theories of the known interactions exhibit the required 
properties. The arguments presented above make these properties
appear far less astonishing predictions; they rather become 
desirable and compelling aspects of these models. While eternal 
inflation seems to be an extremely natural scenario explaining
the formation of (infinitely) many, globally flat universes,
there are many different ways in which one can imagine the
randomness of fundamental constants to be realized in nature.
Superstring theory or M-theory provides an intriguing framework, 
if the conjecture of a multitude of nearly degenerate vacuum 
states is confirmed (but see\cite{KPZ00} for a different view). 
The recently investigated models, in which our observed 
four-dimensional space-time is localized on Dirichlet branes
existing in a higher-dimensional flat or warped 
space\cite{ADD98,RS99}, exhibit enough freedom to allow for 
the scenario advocated here. But other explanations, such 
as the idea that the laws of nature are microscopically 
random\cite{Fo80}, are also possible.

\bigskip
{\em Acknowledgment:}
This manuscript is dedicated to Sergei Matinyan on the occasion
of his seventieth birthday. I wish to thank the students in my
fall semester 2000 class ({\em Physics and the Universe}) for
asking the penetrating questions which inspired this article.
This work was supported in part by a grant from the 
U.~S.~Department of Energy (DE-FG02-96ER40945) and by a Senior 
Scientist Award from the Alexander von Humboldt Foundation.

\end{document}